\documentclass[11pt]{llncs}
\usepackage{graphicx}
\usepackage{amssymb}
\usepackage{epstopdf}
\usepackage{amsmath}
\usepackage[all]{xy}
\newcommand\keywords[1]{\vspace{.1in}\par\noindent{\bf Keywords}: {#1}}
\newcommand\amscode[1]{\vspace{.2in}\par\noindent{\bf 2000 AMS Subject Classification}: {#1}}
\title{A Dimension Reduction Method for Inferring Biochemical Networks}
\titlerunning{<Dimension Reduction for Networks Inference>} 
\author{Gheorghe Craciun\inst{1}\and Casian Pantea\inst{2}\and Grzegorz A. Rempala\inst{3}}
\institute{Department of Mathematics and Department of Biomolecular Chemistry, University of Wisconsin-Madison \email{craciun@math.wisc.edu}
\and Department of Mathematics, University of Wisconsin-Madison \email{pantea@math.wisc.edu}
\and Department of Biostatistics, Medical College of Georgia, Augusta, GA 30912 \email{grempala@mcg.edu}}
\begin{document}
\maketitle
\abstract{We present herein an extension of an algebraic statistical method for inferring biochemical reaction networks from experimental data, proposed recently in \cite{Craciun_Pantea_Rempala}.  This extension allows us to analyze reaction networks that are not necessarily full-dimensional, i.e., the dimension of their stoichiometric space is smaller than the number of species.  Specifically,  we propose to  augment  the original algebraic-statistical algorithm for network inference with  a  pre-processing step that identifies the subspace spanned by the correct reaction vectors, within the space spanned by the species.
This dimension reduction step is based on   principal component analysis of the input data and its relationship with various subspaces generated by sets of candidate reaction vectors.  Simulated examples are provided to illustrate the main ideas involved in implementing this method, and to asses its performance.
}
\amscode{92C40, 92C45, 52B70, 62F}

\section{Introduction}

Modern biological research often involves collecting detailed information on time-dependent chemical concentration data for complex networks or pathways of biochemical interactions \cite{Crampin,Maria}. 
Often, the main purpose of collecting such data is to extract information on the structure of a network of biochemical reactions for which only the identity of the chemical species present in the network is known, with little or no information on the structure of the network of interactions between these species \cite{mar_cal}. 
This problem is of particular interest in the context of molecular and systems biology, and has received a lot of attention in the literature for a long time \cite{bensal,Fay_Balogh,Himmelau_Jones_Bischoff,Hosten,Karnaukhov_etal_2007,Rudakov_1960,Rudakov_1970,Schuster_Hilgetag_Woods_Fell,Vajda_Valko_Yermakova}. 

Two different reaction networks might generate identical (deterministic) mass-action models, making it impossible to discriminate among them, even if we are given experimental data of perfect accuracy and unlimited temporal resolution (this lack of uniqueness is sometimes referred to as the {\it ``fundamental dogma of chemical kinetics"} \cite{Crampin,Erdi_Toth,Epstein}). Necessary and sufficient conditions for two reaction networks to give rise to the same deterministic mass action model are described in \cite{Craciun_Pantea}, where the problem of identifiability of mass action reaction network models was analyzed in detail. The key observation is that, if we think of reactions as vectors, it is possible for different sets of such vectors to span the same positive cones, or at least to span positive cones that have nonempty intersection (see Figure 1 in \cite{Craciun_Pantea} for a simple example).

Often, the experimental measurements for the study of a specific reaction network or pathway are being collected under many different experimental conditions, which affect the values of reaction rate parameters. Moreover, the reactions of interest may not be ``elementary reactions", for which the reaction rates parameters must be constant, but so-called ``overall reactions" which summarize several elementary reaction steps. In that case the reaction rates parameters may reflect the concentrations of biochemical species which have not been included explicitly in the model. Then the reaction rate parameters are {\it not} constant, but rather depend on specific experimental conditions, such as concentrations of enzymes and other intermediate species. Therefore, the estimated vector of reaction rate parameters will not be the same for all experimental conditions, but each specific experimental setting will give rise to one such vector of parameters. However, the set of all these vectors should  span a specific cone, whose extreme rays should identify exactly the set of reactions that gave rise to the data \cite{Craciun_Pantea}. 

Based on these geometric considerations, we proposed a statistical method \cite{Craciun_Pantea_Rempala} which allows us to take advantage of the inherent stochasticity in the data, in order to determine the {\it unique} reaction network that can best account for the results of {\it all} the available experiments pooled together. This idea is related to the notion of {\em algebraic statistical model} (see Chapter 1 in \cite{Pachter_Sturmfels}), and relies on mapping the estimated reaction parameters into an appropriate convex region of the span of reaction vectors of a network, and using the underlying geometry to identify the reactions which are most likely to span that region.   This approach reduces the network identification problem to a statistical inference problem for the parameters of a multinomial distribution, which may then be solved using classical likelihood methods. The description of the computational approaches to analyzing  the geometry of convex regions of interest in our setting  may be found in   \cite{HY08}.

In the method described in \cite{Craciun_Pantea_Rempala} we have made the assumption that the dimension of the span of the ``true" reaction vectors is maximal, i.e., it equals the total number $d$ of species; this assumption allowed us to relate $(i)$ the likelihood of a $d$-dimensional convex cone spanned by a particular set of $d$ reaction vectors with $(ii)$ the number of experimental data points that lie in this cone.

\begin{figure}[!t]\label{cloud3.pdf}
\centering
\includegraphics[width=4.5in]{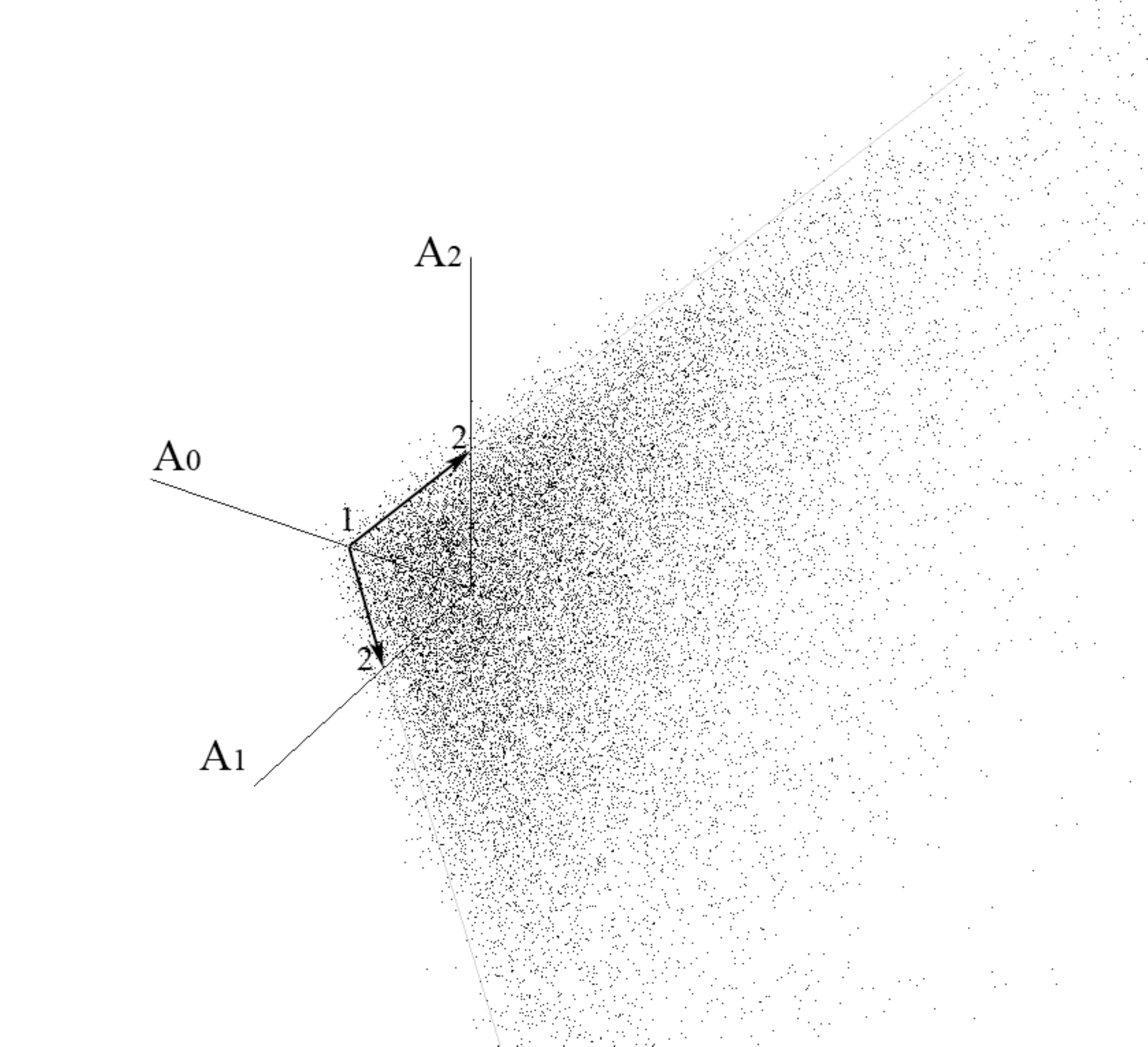} 
 \caption{\small Simulated data points for the network $\{A_0 \to 2A_1,   A_0 \to 2A_2\}$ were obtained by using the procedure described in Section 3. Gaussian noise of mean zero and SD 0.15 was also added. Note that the points remain close to the positive cone spanned by the two reaction vectors. This cone has dimension two, while the space spanned by the species has dimension three.}
\end{figure}

In this paper we extend the applicability of the maximum likelihood method to reaction networks that are not necessarily full-dimensional in the sense explained above. An example of such a network is 
\begin{equation}\label{ex:small}
A_0\to 2A_1,\quad A_0\to 2A_2,
\end{equation}
where we have three species but the reaction vectors $[-1,\ 0,\ 2]$ and $[-1,\ 2,\ 0]$ span a two dimensional cone. In Section 3 of the paper we explain how we can represent the data as geometric coordinates of points in the interior of the cone generated by the true reaction vectors.  A cloud of data points simulated from network (\ref{ex:small}) is depicted in Figure \ref{cloud3.pdf}. Although it is altered by noise, the data set has a flat configuration parallel to the plane of $2A_0-A_1-A_2=0$ determined by the stoichiometric vectors. The new pre-processing algorithm introduced in this paper will recognize the true dimensionality of the data using an eigenvalue analysis approach, and will process the data so the maximum likelihood method in \cite{Craciun_Pantea_Rempala} can be applied.

\section{Maximum Likelihood Method for Inferring Biochemical Reaction Networks}

\bigskip

Given experimental data that originates from a large network of possible reactions, a systematic way of inferring the subset of reactions that are most likely to generate the data was introduced in \cite{Craciun_Pantea_Rempala}. The method uses a statistical model that parametrizes the probabilities that the given reactions are part of the correct subset. The three main assumptions for the model are 

\begin{itemize}
\item[1.] the network is {\em conical}, i.e., all reactions have a common source,
\item[2.] the true reactions generate linearly independent reaction vectors,
\item[3.] the stoichiometric matrix has full rank, i.e., the linear span of the true reaction vectors is the same as the span of the species. 
\end{itemize}

Assumption~1 does not actually limit the scope of this method since, as shown in \cite{Craciun_Pantea}, the network inference can be done one source at a time. On the other hand, the full-dimensionality assumption~3 does not hold for large classes of networks, for example networks that obey one or more linear conservation laws, such as the network (\ref{ex:small}). In this paper we extend the applicability of this maximum likelihood method to these classes of reaction networks by introducing a {\it dimension reduction} pre-processing step. In future work we intend to analyze the case where assumption~2 does not hold as well.

The general setup of the maximum likelihood method in \cite{Craciun_Pantea_Rempala} is as follows. We consider $d$ species and a set of $m$ possible reactions ${\bf R} = \{R_1,\ldots, R_m\}$ involving these species, and we denote by ${\cal R}_d$ the set of all $m\choose{d}$ positive cones spanned by subsets of $d$ reactions in ${\bf R}$. Further we denote by $Cone({\bf R})$ the positive cone generated by the reaction vectors in ${\bf R}$ and by $\{S_1,\ldots, S_n\}$ the partition of $Cone({\bf R})$ into smallest full-dimensional sub-cones obtained by taking all possible intersections of cones in ${\cal R}_d$. The full dimensional cones  $S_1,\ldots, S_n$ are referred to as  {\em building blocks.}

A parameter $\theta_i$, $i=1,\ldots m$ is defined as the conditional probability of the reaction $R_i$ to be involved in generating the data, given that there are $d$ true reactions. These parameters are used to define functions $p_j(\theta_1, \dots ,\theta_m)$, $j=1,\ldots,n$ which represent the probability that a data point is observed in the building block $S_j,$ conditioned on the existence of exactly $d$ true reactions. The construction of the $p_j$'s uses appropriate volume measurements to asses the probability that a point lies inside a certain cone. In doing so, the method disregards the zero volume (degenerate) cones and it is therefore not appropriate for data generated from a network for which the true reactions do not span a full dimensional cone. See   \cite[Section 2]{Craciun_Pantea_Rempala} for detailed explanation of the geometric methods in the construction of $p_j.$ A likelihood function is constructed using the functions $p_j$ and also involves counting the data points within each building block $S_1,\ldots, S_n$. We then infer the true reactions from the values of $\theta_i$ that maximize the likelihood function. 

The algorithm was implemented using Matlab \cite{BioNetID} and  has been made publicly available at {\tt http://www.math.wisc.edu/$\sim$pantea}. To illustrate the implementation, consider the set of reactions in (\ref{CRN2}), and data generated from all these reactions except for $A_0\to A_2$, using independent rate constants drawn from a $Gamma(1.5, 1)$ distribution (see also the discussion in \cite[Example 7]{Craciun_Pantea_Rempala}).

\begin{equation}\label{CRN2}
\xy ;<1pc, 0pc>:
\POS(0,0)*+{A_{0}}
\ar@<.3 ex> @{->}        +(-3,-3)*+{A_{1}+A_2}
\ar@<.3 ex> @{->}        +(-3,3)*+{A_{1}}
\ar@<.3 ex> @{->}        +(0,3)*+{2A_3}
\ar@<.3 ex> @{->}        +(3,3)*+{A_2}
\ar@<.3 ex> @{->}        +(3,-3)*+{A_2+A_3}
\endxy
\end{equation}

Once the likelihood function is constructed, the algorithm runs an optimization procedure for the maximum of the likelihood, repeated $2^{m-1}$ times  (e.g., 16 times for network (\ref{CRN2})), with restarts from random initial values of $\theta$. A list of (local) minima  is created, and the algorithm reports the value of $\theta$ that realized the smallest local minimum,  together with the percentage of time the algorithm ends up at that particular point (i.e., the success rate). An example of such a report for ${\bf R} = \{ A_0 \to 2A_3, A_0 \to A_1, A_0 \to A_1+A_2, A_0 \to A_2+A_3, A_0 \to A_2 \}$ is

\vspace{.2cm}
\noindent \texttt{Minimum of negative log-likelihood: 6.94\\  Theta:\\
1\hspace{2cm}1\hspace{2cm}1\hspace{2cm}1\hspace{2cm}0.\\
Hits: 16 out of 16, 100\%.}
\vspace{.2cm}

\noindent This shows that the algorithm selects the correct four reactions as the ones generating the data by assigning them probability 1, and discards the last (incorrect) reaction $A_0 \to A_2$ by assigning it probability 0.

\section{Numerical Example with Simulated Input Data}

We describe our approach for dimension reduction by analyzing in detail a modified version of reaction network (\ref{CRN2}).  Let  $A_{i}$, $i = 0,1,2,3$, denote four chemical species and consider the following augmented network of possible reactions: 

\begin{equation}\label{CRN0}
\xy ;<1pc, 0pc>:
\POS(0,0)*+{A_{0}}
\ar@<.3 ex> @{->}        +(-3,-3)*+{A_{1}+A_2}
\ar@<.3 ex> @{->}        +(-3,3)*+{A_{1}}
\ar@<.3 ex> @{->}        +(0,3)*+{2A_3}
\ar@<.3 ex> @{->}        +(3,3)*+{A_2}
\ar@<.3 ex> @{->}        +(3,-3)*+{A_2+A_3}
\ar@<.3 ex> @{->}        +(0,-3)*+{2A_2}
\endxy
\end{equation}
where we suppose that the true reactions  that generate the data are 

\begin{equation}\label{true}
A_0\to A_1+A_2,\ \ A_0\to A_2+A_3,\text{ and } A_0\to 2A_3.
\end{equation}

In this setting we would like to test whether  our algorithm would correctly  identify the three reactions (\ref{true}) out of the network (\ref{CRN0}) as being responsible for generating the data.

Note that unlike  \cite[Example 8]{Craciun_Pantea_Rempala}, where the set of true reactions was taken to be 

\begin{equation}\label{true2}
A_0\to A_1,\ A_0\to A_1+A_2,\ A_0\to A_2+A_3,\text{ and } A_0\to 2A_3,
\end{equation} 

\noindent the set of true reactions (\ref{true}) spans a three-dimensional (instead of four-dimensional) space. Since the original  maximum likelihood algorithm as proposed  in \cite{Craciun_Pantea_Rempala} required full-dimensionality of the set of true reactions,  it cannot be applied directly in our case. A pre-processing step is required to determine the correct dimensionality of the data.

\bigskip
 
\noindent \textbf{Data generation.} We  use the true reactions  (\ref{true}) above to simulate ``experimentally measured" data and to test  whether our dimension reduction procedure, followed by the likelihood-based algorithm from \cite{Craciun_Pantea_Rempala}, is able to identify the true reactions from the the possible six reactions of network (\ref{CRN0}). 

The deterministic dynamics of the chemical reaction network (\ref{CRN0}) is governed by  linear differential equations of the form 

\begin{equation}\label{eq:rre} 
dA_i/dt=\gamma_i A_0\qquad i=0,\ldots,3.
\end{equation} 
where the parameters $\gamma_i$, $i=0\ldots 3$ are linear combinations of the rate constants, given by the stoichiometry of the true network.

To obtain the simulated measurements of the concentrations of $A_0,\ A_1,\ A_2$ and  $A_3$  we independently draw the three rate parameters of the true reactions from a gamma distribution $Gamma(\alpha,\lambda)$ with parameters $\alpha=1.5$ and $\lambda=1$ and use a standard Matlab ODE solver to generate trajectories of the resulting mass-action system. We then fit the simulated trajectories to system (\ref{eq:rre}) using the Matlab function \texttt{lsqcurvefit} and obtain a single resulting data point $K=(\hat{\gamma}_1,\hat{\gamma}_2,\hat{\gamma}_3,\hat{\gamma}_4).$  This procedure is repeated to generate 20 data points.

The data points $K_i$, $i=1,\ldots 20,$  can be viewed geometrically as points in the coordinate system given by the species $[A_0, A_1, A_2, A_3]$. This interpretation is not related to any choice of the reactions. However, if we disregard the fitting error, each data point lies inside the three dimensional open convex cone generated by the true reaction vectors. Moreover, as explained in \cite{Craciun_Pantea}, the corresponding conical coordinates of  $K_i$  are precisely the estimates of the true rate constants.

\bigskip 

\noindent \textbf{Dimension reduction.}  Although our data is given as four dimensional points, it is in fact generated by a reaction network whose stoichiometry is three dimensional. Therefore the data points are confined to a three-dimensional subspace of the canonical species space $[A_0, A_1, A_2, A_3]$, or are very close to a three-dimensional subspace, if one accounts for the error that occurs in the estimation.

We introduce a new first step to our algorithm, which consists of recovering the true dimensionality of the data. To this end we use a standard eigenvalue method, namely we look for small  eigenvalues of the $DD'$ matrix, where $D$ is the $4\times 20$ matrix of data points. Here ``small"  eigenvalues correspond in fact to zero  eigenvalues, altered by estimation error. The corresponding eigenvectors impose linear constraints on the data, and consequently on the true reaction vectors. We can therefore discard the reaction vectors that do not satisfy the aforementioned constraints. 

For the network (\ref{CRN0}) with the true reactions (\ref{true}) the first output of the algorithm is

\vspace{.3cm}

\noindent {\tt Rays :} 
\hspace{-.2cm}$\begin{matrix}
    &\mathtt{-1}     &\mathtt{1}     &\mathtt 0     &\mathtt 0\\
    &\mathtt{-1}     &\mathtt{0}     &\mathtt 2     &\mathtt 0\\
     &\mathtt{-1}     &\mathtt{0}     &\mathtt 0     &\mathtt 2\\ 
     &\mathtt{-1}     &\mathtt{0}     &\mathtt 1     &\mathtt 0\\
    &\mathtt{-1}     &\mathtt{0}     &\mathtt 1     &\mathtt 1\\
    &\mathtt{-1}     &\mathtt 1     &\mathtt 1     &\mathtt 0 \\
\end{matrix}\quad $ 
{\tt    Discarded: }
\hspace{-.2cm}$\begin{matrix}
    &\mathtt{-1}     &\mathtt{1}     &\mathtt 0     &\mathtt 0\\
    &\mathtt{-1}     &\mathtt{0}     &\mathtt 1     &\mathtt 0\\
\end{matrix}\quad $
{\tt Possible:}
\hspace{-.2cm}$\begin{matrix}
    &\mathtt{-1}     &\mathtt{0}     &\mathtt 2     &\mathtt 0\\
    &\mathtt{-1}     &\mathtt{0}     &\mathtt 0     &\mathtt 2\\
    &\mathtt{-1}     &\mathtt{0}     &\mathtt 1     &\mathtt 1\\
    &\mathtt{-1}     &\mathtt 1     &\mathtt 1     &\mathtt 0 \\
\end{matrix}$

\vspace{.2cm}

\bigskip

The algorithm has correctly concluded that the data points should belong to the hyperplane $2A_0+A_1+A_2+A_3=0$ and that the true reactions are among the four whose reaction vectors are included in this hyperplane, namely
$$A_0\to 2A_2,\ A_0\to 2A_3,\  A_0\to A_2+A_3,\ A_0\to A_1+A_2.$$

The dimension reduction is completed by projecting our data onto the true stoichiometric subspace above. The algorithm then computes an orthonormal basis of this subspace and rewrites the data in the new coordinates. The resulting data matrix is of dimension $3\times 20$ and the maximum likelihood algorithm in \cite{Craciun_Pantea_Rempala} can now be applied and produces the following output: 

\vspace{.2cm}
\noindent \texttt{Minimum of negative log-likelihood: 0\\  Theta:\\
0\hspace{2cm}1\hspace{2cm}1\hspace{2cm}1\\
Hits: 8 out of 8, 100\%.}
\vspace{.2cm}

\noindent which  refers to the reaction vectors labeled ``possible"  and correctly concludes that the true reactions are the ones corresponding to the last three reaction vectors from the ``possible" list. 

Note that the new coordinates of the data points projected onto the subspace of the true stoichiometry do not have an immediate geometric interpretation, as was the case for the the four coordinates $(\gamma_0, \gamma_1,\gamma_2,\gamma_3).$ However, the geometric arguments that make the  likelihood method work can clearly be applied in the projected configuration.  By choosing an orthonormal basis for the true stoichiometric subspace, we are ensuring that angles and lengths are consistent between the two coordinate systems.

\bigskip 

\begin{figure}[!t]\label{Rplot}
\centering
\includegraphics[width=5in]{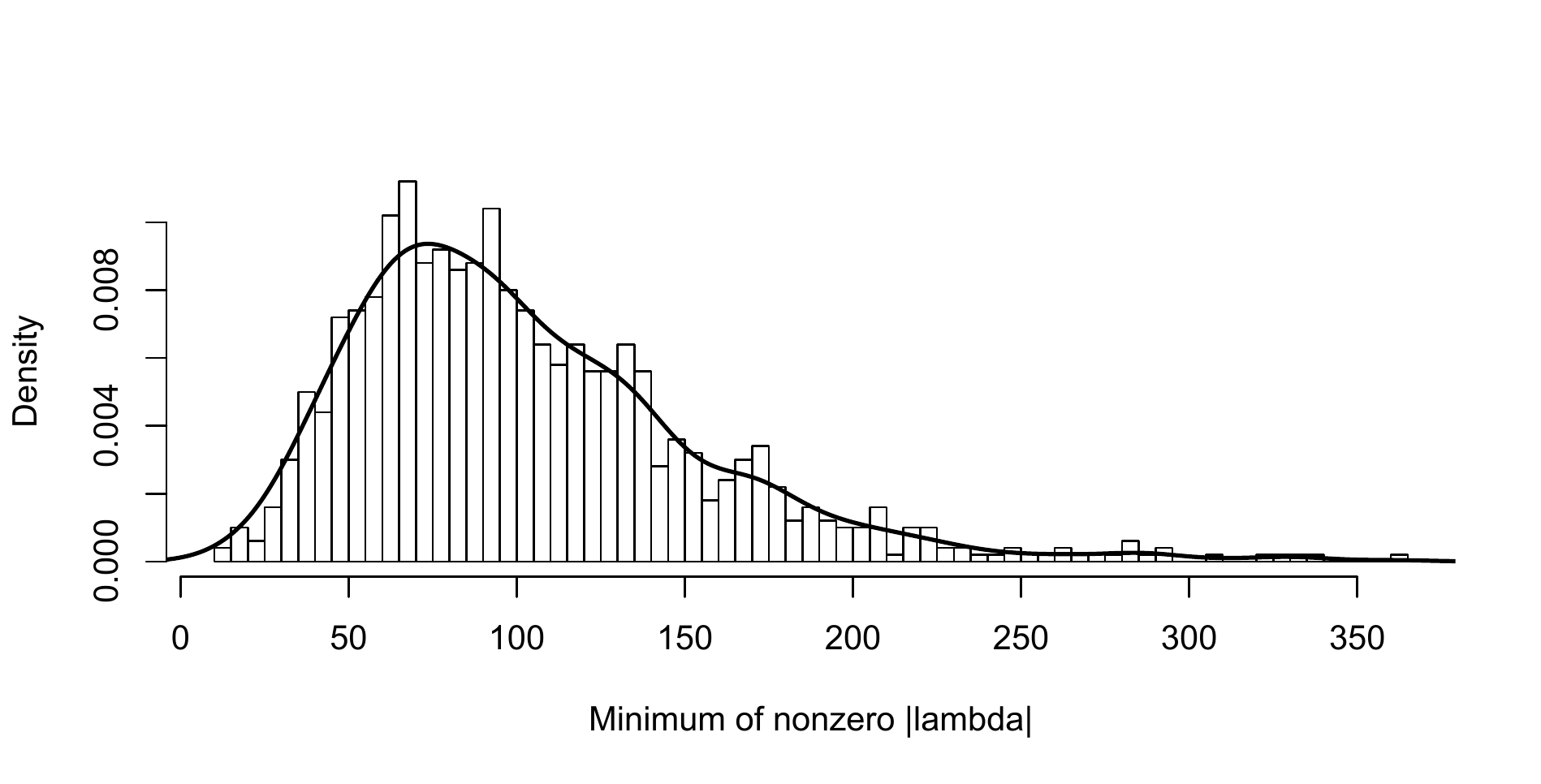} 
 \caption{\small Density plot for the empirical distribution of the smallest nonzero absolute value $\lambda$ among the eigenvalues of $DD'$. The reactions that generate $D$ are  $\{A_0 \to 2A_2,   A_0 \to A_1+A_2.\}$}
\end{figure}

\noindent \textbf{Threshold for eigenvalues.} Our dimension reduction step is dependent upon isolating small eigenvalues, which would be zero had it not been for the estimating error. Therefore we need a threshold value to separate the eigenvalues that are small because they correspond to zero eigenvalues perturbed by error.  To obtain this threshold, we use a Monte Carlo simulation to construct the empirical distribution of the smallest nonzero absolute value of an eigenvalue of $DD'$. The data matrix $D$ has as many points as the real data (20 in our case) and is generated  from all possible $2^6-1=63$ choices of the true reactions out of the possible 6 reactions in network (\ref{CRN0}), without any noise added. For each such combination of reactions a sample of 1000 $D$ matrices was generated using independent  $Gamma(1.5,\ 1)$ distributed rate constants. An illustrative histogram of the typical resulting empirical density of the smallest eigenvalue in absolute value is given in  Figure \ref{Rplot}. This is just one of the 63 cases, the case where the true reactions are taken to be $A_0 \to 2A_2$ and   $A_0 \to A_1+A_2.$

We merge all the 63 frequencies to obtain a unique distribution for the network of possible reactions (\ref{CRN0}) and we use a standard quantile of this distribution to define the threshold value. In example (\ref{CRN0}), the 1\% quantile gives a value of the threshold of  0.7.


\bigskip

\noindent{\bf More numerical results.} One could further test the algorithm by choosing the set of true reaction vectors set to be even lower dimensional, e.g., of dimension two (for example $A_0\to 2A_2,\ A_0\to 2A_3,\ A_0\to A_2+A_3$)  or one (for example, $A_0\to A_1$). For the particular network considered in the  tests we conducted, both  of these dimension reductions were  performed correctly by our algorithm pre-processing step  and moreover the subsequent maximum likelihood step always had 100\% success rate. This is perhaps not surprising because for small dimensional problems of this type the maximum likelihood algorithm is expected to behave very well.

\bigskip

\begin{figure}\label{noise} 
\centering
\includegraphics[width=3in]{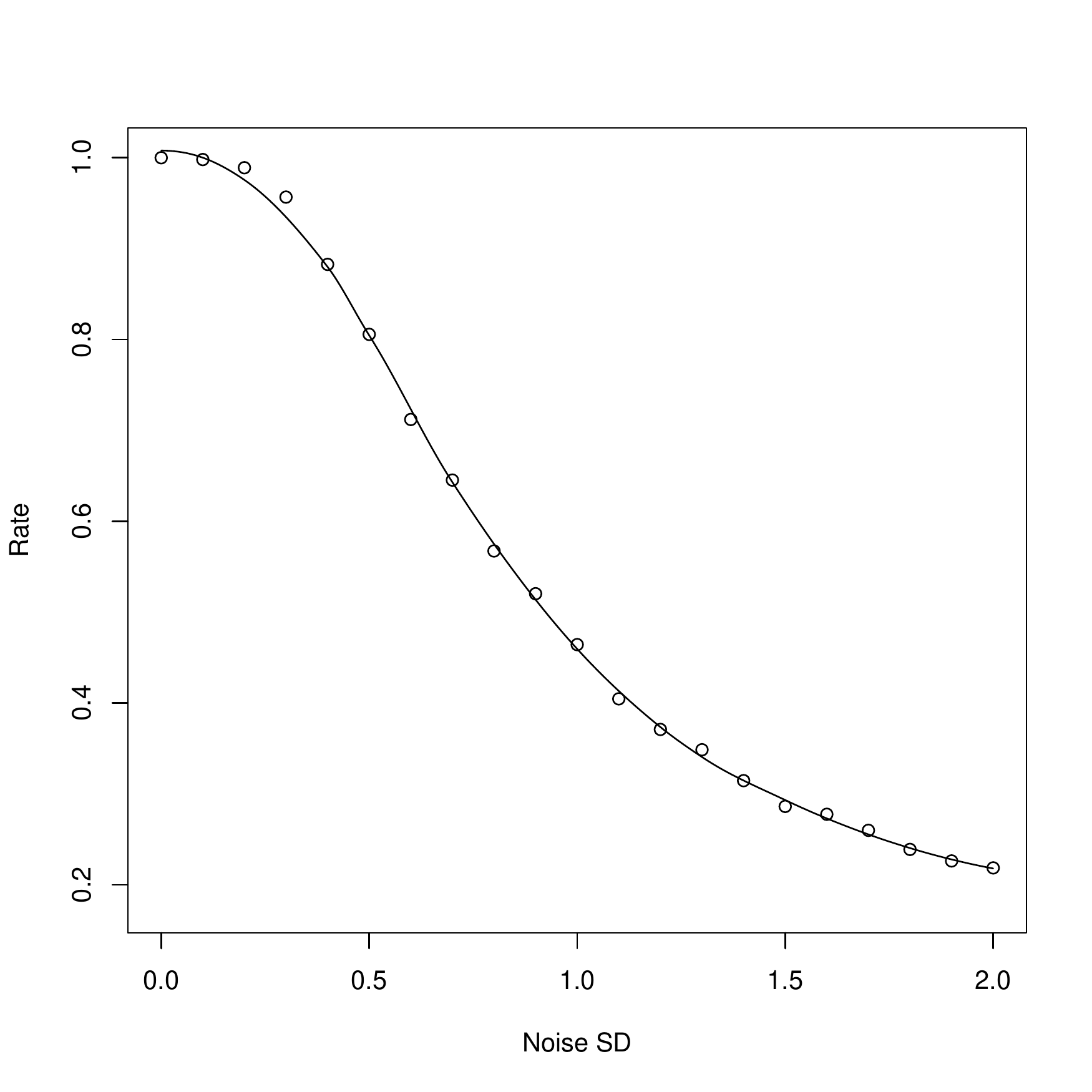} 
\caption{\small The rate of recovery of  the correct reactions (\ref{true2}) out of the network \eqref{CRN0} as a function of the size of the Gaussian noise added to the estimated parameters $[\hat{\gamma}_1,\hat{\gamma}_2,\hat{\gamma}_3,\hat{\gamma}_4].$ The recovery rate is over 90\% for the Gaussian  noise with 0.5 SD and about 50\% for  1 SD.}
\end{figure}

\noindent{\bf The effect of noise.} In \cite{Craciun_Pantea_Rempala} the effect of the estimating error on the effectiveness of the maximum likelihood algorithm was studied using simulations. Namely, zero-mean Gaussian noise with varying variance was added to a set of data points $K_i$,  $i=1\ldots 20$ corresponding to the reaction network (\ref{CRN0}) and true reactions (\ref{true2}). This was repeated in batches of 3000 for increasing values of the standard deviation of the added noise and the maximum likelihood algorithm for network discovery was performed. The result of this experiment is presented in Figure \ref{noise}. As  seen from the figure with small to moderate random noise added to the values of the reaction constants the likelihood method is still able to recover the correct set of reactions at remarkably high rate. 

\begin{figure}\label{noise3}
\centering
\includegraphics[width=3in]{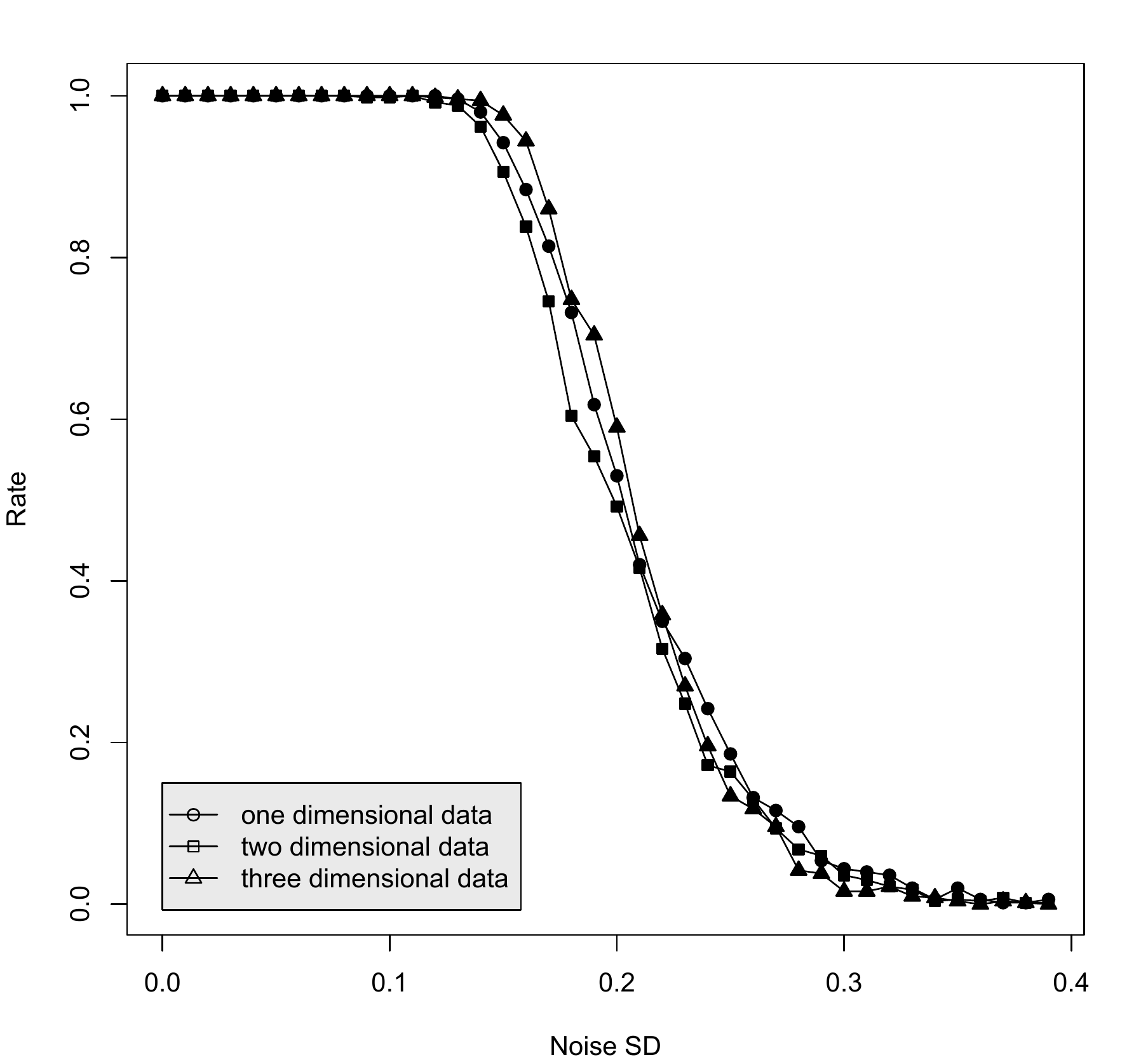} 
\caption{\small The rate of recovery of the correct stoichiometric space for the augmented network \eqref{CRN0} as a function of the size of the Gaussian noise added in batches of 500 to the 20 data points. The three different curves correspond to different dimensions of the true data. } 
\end{figure}

We take a similar approach to investigate how the dimension reduction step is affected by noise. Just as before, we add zero-mean normal noise to the matrix $D$ of data points in batches of 500 for 50 increasing values  of the noise standard deviation, from  0 to .5. We consider three qualitatively different cases for the true network that generated the data, according to the dimension of the corresponding stoichiometry.  More precisely we chose the one-dimensional reaction network $A_0\to 2A_2$, the two dimensional network   $A_0\to 2A_2,\ A_0\to 2A_3,\ A_0\to A_2+A_3$ and the three dimensional network  $A_0\to 2A_3,\  A_0\to A_2+A_3,\   A_0\to A_1+A_2.$  The recovery rate of the true underlying stoichiometric space is presented in Figure \ref{noise3} and has a decreasing shape  similar to the one of Figure \ref{noise}. Note, however, that the dimension recovery rate descends much faster to zero, being much more sensible to noise. In other words, even if the noise does not move  the data points from the inside to the outside of the full dimensional cones generated by the reaction vectors, it can be sufficient  to ``thicken" the distribution of the data points making it look full dimensional and thus misleading the algorithm into choosing a full dimensional stoichiometry.

\bigskip

\section{Summary and Discussion}

A likelihood-based algebraic statistical model for inferring biochemical reaction networks was proposed recently in  \cite{Craciun_Pantea_Rempala}, in order to overcome the unidentifiability problem for chemical networks given by a set of reaction rate equations. Indeed, as  illustrated in the earlier work of some of the authors \cite{Craciun_Pantea}, in the usual deterministic sense such networks are in general unidentifiable, i.e., different chemical reaction networks may give rise to exactly the same reaction rate equations.
The  attractiveness  of the algebraic statistical method was in its ability to take advantage of  the  algebraic and geometric structure  of the network rather than merely the observed experimental values of the network species, as is  commonly the case in  network inference models based on  graphical methods, like e.g. Bayesian or probabilistic boolean networks (cf. \cite{rich}). 
The original algebraic model described in \cite{Craciun_Pantea_Rempala} made the  simplifying assumption that the stoichiometric subspace of the true network was fully-dimensional, i.e., its dimension equaled the number of species. In the current work we remove this restriction via a pre-processing algorithm that identifies the 
true dimensionality of the data using an eigenvalue analysis approach; then, once we identified the correct network span, we use a restricted version of the same algebraic statistical model to identify the correct network. 
We present a small study based on simulated data, where the true dimension of a network was 1-3 units lower than the number of species, and where our dimension reduction method worked as intended. Overall, in the examples provided as well as in other similar numerical experiments not reported here, the algorithm with pre-processing for dimension reduction performed very well, being able to recover  the true network structure almost perfectly as long as the level of noise present in the data was not too large. 

More challenging studies are needed in order to determine the applicability of this method for large biochemical networks and for true experimental data. Also, in future work, we intend to address the case where the assumption that the reaction vectors are linearly independent does not necessarily hold true.

\bigskip

\noindent {\bf Acknowledgments.} 
This research was partially sponsored by the  ``Focused Research Group" grants NSF--DMS 0840695 (Rempala) and  NSF--DMS 0553687 (Craciun) as well as by the NIH grant 1R01DE019243-01 (Rempala) and the NIH grant 1R01GM086881-01 (Craciun).


\begin{thebibliography}{}

\bibitem{bensal} M.~Bansal,  V.~Belcastro,  A.~Ambesi-Impiombato and  D.~di Bernardo, 
 How to infer gene networks from expression profiles. 
Molecular Systems Biology 3:78 (2007)
http://www.molecularsystemsbiology.com DOI:10.1038/msb4100120

\bibitem{BioNetID} G.~Craciun, C.~Pantea, G.~Rempala, 
BioNetID -- Biochemical Networks Identification Software (2009) http://scb.mcg.edu/software
or http://www.math.wisc.edu/$\sim$pantea  


\bibitem{Craciun_Pantea_Rempala} G.~Craciun, C.~Pantea, G.~Rempala,
Algebraic Methods for Inferring Biochemical Networks: a Maximum Likelihood Approach, 2008, arXiv:0810.0561v2


\bibitem{Craciun_Pantea} G.~Craciun, C.~Pantea,
Identifiability of chemical reaction networks,
 Journal of Mathematical Chemistry 44:1, 244-259, 2008.


\bibitem{Crampin} E.J.~Crampin, S. Schnell, and P.E. McSharry,
Mathematical and computational techniques to deduce complex biochemical reaction mechanisms,
 Prog. Biophys. Mol. Biol. 86 (2004) 177.


\bibitem{Erdi_Toth} P.~Erdi and J.~Toth,
 Mathematical Models of Chemical Reactions: Theory and Applications of Deterministic and Stochastic Models, (Princeton University Press, 1989)



\bibitem{Epstein} I.R.~Epstein and J.A.~Pojman,
 An Introduction to Nonlinear Chemical Dynamics: Oscillations, Waves, Patterns, and Chaos,
Oxford University Press, 2002.


\bibitem{Fay_Balogh} L.~Fay and A.~Balogh,
Determination of reaction order and rate constants on the basis of the parameter estimation of differential equations,
 Acta Chim. Acad. Sci. Hun. 57:4 (1968) 391.



\bibitem{Himmelau_Jones_Bischoff} D.M.~Himmeblau, C.R.~Jones and K.B.~Bischoff,
Determination of rate constants for complex kinetic models,
 Ind. Eng. Chem. Fundam. 6:4 (1967) 539.

\bibitem{Hosten} L.H.~Hosten,
A comparative study of short cut procedures for
parameter estimation in differential equations,
 Computers and Chemical Engineering 3 (1979) 117.

\bibitem{HY08} Huggins, P and  Yoshida, R. (2008)
First steps toward the geometry of cophylogeny. Manuscript, available at \texttt{oai:arXiv.org:0809.1908}.


\bibitem{Karnaukhov_etal_2007} A.~Karnaukhov, E.~Karnaukhova and J.~Williamson,
Numerical Matrices Method for Nonlinear System Identification and Description of Dynamics of Biochemical Reaction Networks,
 Biophys. J. 92 (2007) 3459.


\bibitem{Maria} G.~Maria, 
A review of algorithms and trends in kinetic model identification
for chemical and biochemical systems,
 Chem. Biochem. Eng. Q. 18:3 (2004) 195.

\bibitem{mar_cal} A.~Margolini and  A.~Califano,
\newblock Theory and Limitations of Genetic Networks Inference from Microarray Data
\newblock Annals N.Y. Acad Sci. 1115: 51--72  (2007). 
  \newblock www.interscience.wiley.com DOI:10.1002/sim.3017.


\bibitem{Pachter_Sturmfels} L.~Pachter, B.~Sturmfels,
 Algebraic Statistics for Computational Biology,
Cambridge University Press, 2005.


 \bibitem{rich}
 T.~Richardson and P.~Spirtes (2002). Ancestral graph Markov Models. 
Annals of Statistics 30:962--1030




\bibitem{Rudakov_1960}
E.~Rudakov,
Differential methods of determination of rate constants of noncomplicated chemical reactions,
 Kinetics and Catalysis 1 (1960) 177.

\bibitem{Rudakov_1970}
E.~Rudakov,
Determination of rate constants. Method of support function,
Kinetics and Catalysis 11 (1970) 228.


\bibitem{Schuster_Hilgetag_Woods_Fell}
S.~Schuster, C.~Hilgetag, J.H.~Woods and D.A.~ Fell,
Reaction routes in biochemical reaction systems: algebraic properties, validated calculation procedure and example from nucleotide metabolism,
 J. Math. Biol. 45 (2002) 153.



\bibitem{Vajda_Valko_Yermakova}
S.~Vajda, P.~Valko and A. Yermakova,
A direct-indirect procedure for estimating kinetic parameters,
 Computers and Chemical Engineering 10 (1986) 49.




\end{thebibliography}
\end{document}